\documentclass[finallayout,an,fleqn]{w-art}
\usepackage{times}
\usepackage{w-thm}

\usepackage[]{graphicx}
\chardef\bslash=`\\ 

\def\ga{\mathrel{\hbox{\rlap{\hbox{\lower4pt\hbox{$\sim$}}}\hbox{$>$}}}}
\def\la{\mathrel{\hbox{\rlap{\hbox{\lower4pt\hbox{$\sim$}}}\hbox{$<$}}}}
\def\arcsec{\hbox{$^{\prime\prime}$}}

\newcommand{\kms}{~km\,s$^{-1}$} 
\newcommand{\ergs}{~erg\,s$^{-1}$\,cm$^{-2}$\,sr$^{-1}$}

\newcommand{\msol}{\hbox{$M_\odot$}}
\newcommand{\lsol}{\hbox{$L_\odot$}}

\newcommand{\oh}{OH(1720 MHz)~}
\newcommand{\hone}{2.12~$\mu$m H$_2$ 1--0 S(1)\ }

\begin{document}

\DOIsuffix{theDOIsuffix}
\Volume{324}
\Issue{S1}
\Copyrightissue{S1}
\Month{01}
\Year{2003}
\pagespan{1}{4}
\Receiveddate{}
\Reviseddate{}
\Accepteddate{}
\Dateposted{}
\keywords{shock waves, molecular clouds, supernova remnants, star
formation, Tornado nebula, G357.7--0.1, Eye of Tornado, G357.63--0.06}
\subjclass[pacs]{04A25}



\title[Tornado and its Eye]{Molecular Line Observations of the
  Tornado Nebula and its Eye}

\author[Lazendic]{J. Lazendic\footnote{Corresponding
     author: e-mail: {\sf jlazendic@cfa.harvard.edu}}\inst{1,5,6}}
     \address[\inst{1}]{Harvard-Smithsonian CfA, 60 Garden street,
     Cambridge MA 02138, USA}
\author[Burton]{M. Burton\inst{2}}
\address[\inst{2}]{School of Physics, University of New South Wales,
Sydney NSW 2052, Australia}
\author[Yusef-Zadeh]{F. Yusef-Zadeh\inst{3}}
\address[\inst{3}]{Department of Physics and Astronomy, Northwestern
University, Evanston, IL 60208, USA}
\author[Wardle]{M. Wardle\inst{4}}
\address[\inst{4}]{Department of Physics, Macquarie University, NSW 2109, Australia}
\author[Green]{A. Green\inst{5}}
\address[\inst{5}]{School of Physics, University of Sydney, Sydney NSW 2006, Australia}
\author[Whiteoak]{J. Whiteoak\inst{6}}
\address[\inst{6}]{Australia Telescope National Facility}

\begin{abstract}

 We present millimetre and NIR molecular-line observations of the
 Tornado nebula and its Eye.  The observations were motivated by the
 presence of OH(1720 MHz) maser emission towards the nebula, believed
 to be an indicator of interaction between a supernova remnant and a
 molecular cloud.   We found that the  distribution of molecular gas
 around the Tornado complements its radio morphology, implying that
 the nebula's appearance has been influenced by the structure of the
 surrounding molecular gas.  Our NIR H$_2$ observations revealed the
 presence  of shocked molecular gas at the location where the nebula is
 expanding into the surrounding molecular cloud.

 It has been suggested that the Eye of the Tornado is related to the
 nebula on the basis of  their apparent proximity.  Our NIR and
 millimetre-line observations show that the two objects are not
 spatially related. Br\,$\gamma$  line emission, in conjuction with
 IR data at longer wavelengths and high-resolution radio continuum
 observations, suggests that the Eye is a massive protostellar source 
 deeply embedded within a dense molecular core. 

\end{abstract}

\maketitle                   


\section{Introduction}

The Tornado nebula (G357.7--0.1)  is a peculiar radio source located
towards the Galactic Centre region. It has been classified as a
supernova remnant (SNR) due to its steep radio spectrum and linear
polarization (e.g., Kundu et al. 1974, Caswell et al. 1980, Shaver et
al. 1985a), but its unique morphology has led to other interpretations
(e.g.,  an accretion powered nebula, Becker \& Helfand 1985).  The Eye
of the Tornado (G357.63--0.06) is a compact radio source located
$30\arcsec$ from the emission peak of the nebula. It was initially
thought to be responsible for the formation of the nebula
(e.g., through mass ejection from a pulsar or accreting binary
system), but was instead found to have a flat radio spectrum and was
suggested to be an H{\sc ii} region (Shaver et al. 1985b). 

\section{Tornado}

Frail et al. (1996) found a single \oh maser at the northwestern tip
of the Tornado (see Figure 1). When not accompanied by maser emission
from the other  three OH ground--state transitions at 1612, 1665 and
1667 MHz, the detection of this maser has been recognized as a signature of
SNR/molecular cloud interactions (see Koralesky et al. 1998 and
references therein). Its presence may support the classification of the
nebula as an SNR. The maser has a velocity of $-12.4$\kms,
implying  a distance of 11.8 kpc to the nebula and placing the Tornado
behind the Galactic Centre. 

\begin{figure}[htb]
\includegraphics[height=7cm]{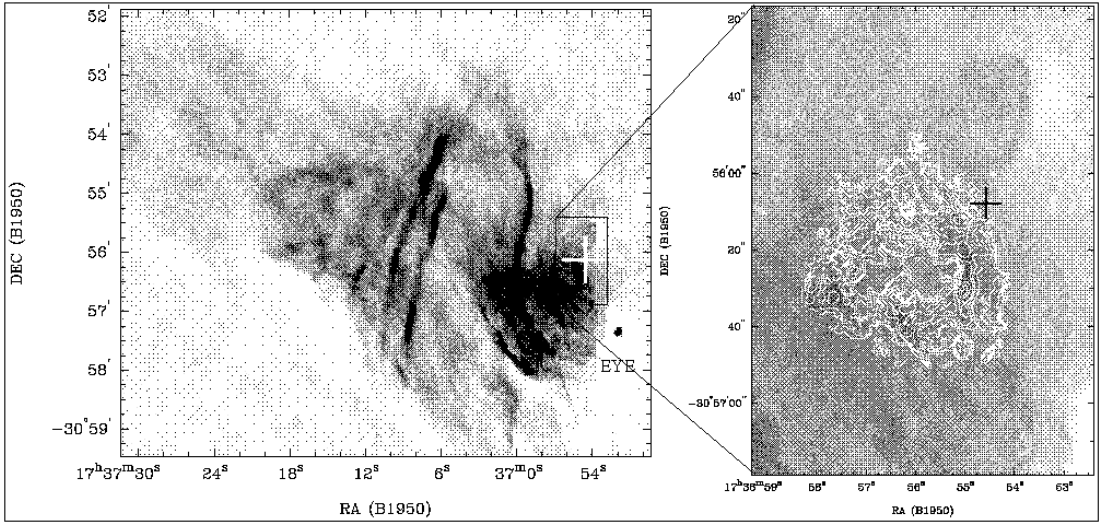}
\caption{{\em (left)} A 20 cm VLA image of the Tornado Nebula. The
square marks the part of the nebula covered by the UNSWIRF observations.
{\em (right)} Contours of H$_2$\ line emission superimposed on a
greyscale 20 cm radio continuum image of the northwestern part of the
Tornado. The contour levels are: 1.6, 4.7, 6.3, 7.9, 9.5, 11.1,
12.6, $14.2 \times 10^{-5}$\ergs. The white and black 
crosses  mark the location of the \oh maser.}
\end{figure}

Using the University of New South Wales Fabry-Perot narrow-band tunable
filter (UNSWIRF), we detected \hone emission
towards the \oh maser in the Tornado nebula (see Figure 1). The
correlation of the emission peaks in the radio continuum and H$_2$\
images suggest that the H$_2$\ emission originates from an expansion of
 a shock wave and is most probably shock excited, as found in other
SNRs associated with the \oh  maser (e.g., Lazendic et al. 2002a,b). The
\oh maser is located at the western edge of the H$_2$\ emission, which is
more sharply defined than the rest of the ring, probably delineating
the leading edge of the shock front.

Molecular transitions at millimetre wavelengths were also detected at
the maser velocity of $-12$\kms\ using the 15-m Swedish-ESO Submillimeter
Telescope (SEST). Emission from molecular species other
than $^{12}$CO\ and $^{13}$CO, e.g., HCO$^+$, HCN and H$_2$CO, was found to be
very weak (see Lazendic et al. 2003 for more details). 
Molecular gas associated with the \oh maser and H$_2$\
emission is optically thick, cold ($\sim$7 K) and dense
($\sim 10^{5}$\,cm$^{-3}$).  This density is in agreement with the
requirements for the \oh maser production in the post-shock gas behind
the SNR shock front (Lockett et al. 1999), but the temperature is much lower than
that expected in the post-shock gas in which the maser is created (50
-- 125\,K). However, since the cloud is optically thick, our CO observations
 are probing only the envelope of the cloud.  Observations of
 more optically thin transitions of $^{12}$CO and $^{13}$CO are needed to examine 
the whole cloud temperature.

\begin{figure}[htb]
\includegraphics[height=10cm]{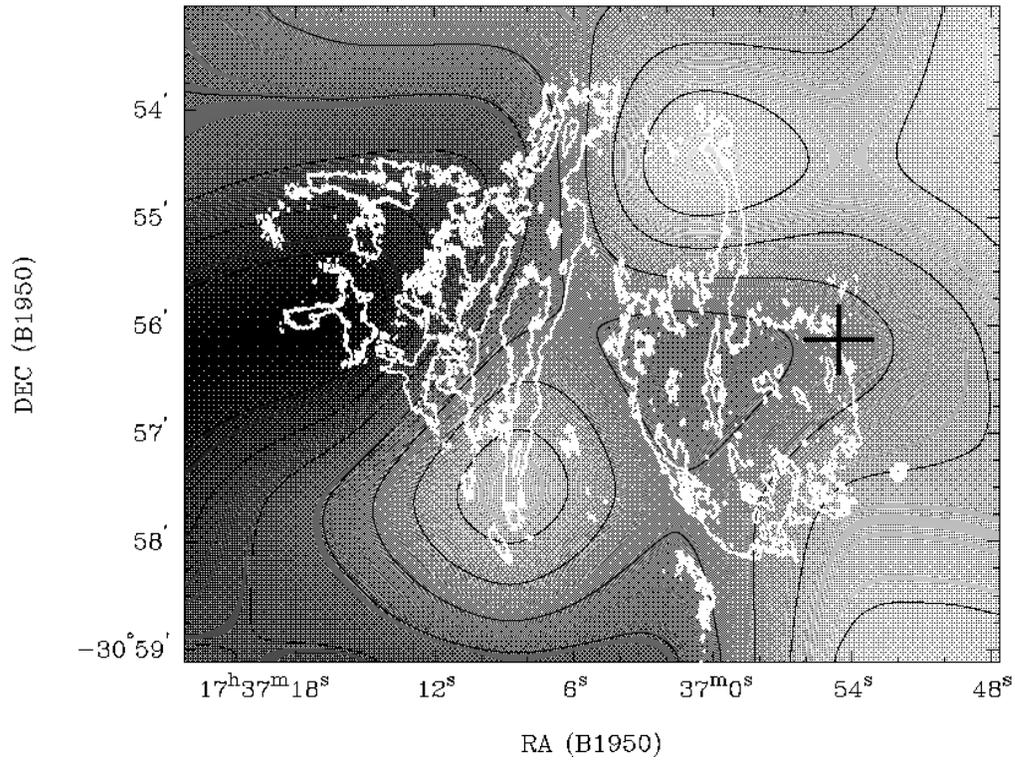}
\caption{An image of $^{13}$CO  1--0 emission (shown in greyscale and black
contours) integrated between $-16$ and $-9$\kms, overlaid with
contours of the 20 cm radio continuum emission. The $^{13}$CO\ contour
levels are: 3.8, 5.7, 7.6, 9.5, 11.4, 13.3, 15.2 and 17.1 K\kms. The
lightest colours represent the weakest emission. The 20 cm contour levels are: 2.
7, 8.1, 13.5, 18.9, $24.3 \times 10^{-2}$\,Jy\,beam$^{-1}$. The cross  marks the location
of the \oh maser.}
\end{figure}

The structure of the associated molecular gas complements the
radio morphology of the Tornado nebula (see Figure 2), implying
that the distribution of the surrounding medium has influenced the nebula's
unusual appearance. In particular,  two minima in the molecular 
gas distribution, located  symmetrically 
on each side of the nebula, coincide with large
arc-like filaments in the nebula and point to locations where
the shock could perhaps expand more readily than in the other regions.

\section{Eye of Tornado}

\begin{figure}[h]
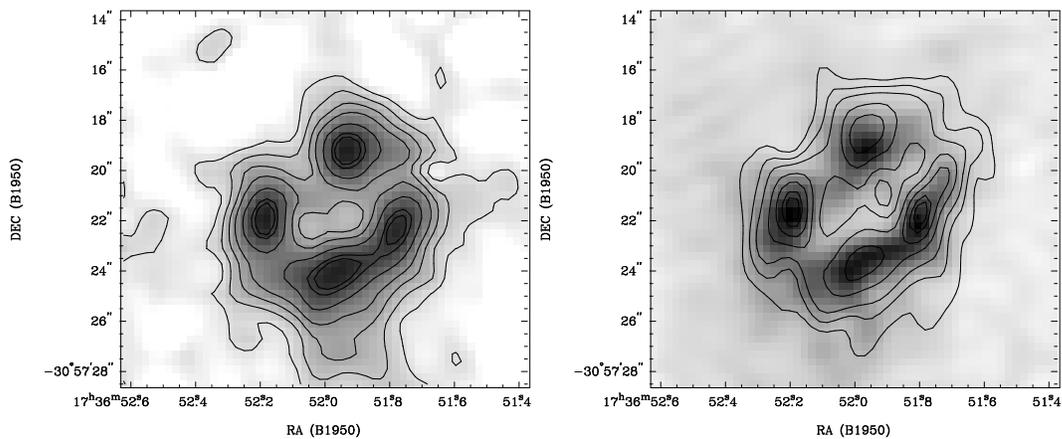

\includegraphics[height=5.7cm]{fig3a.ps}
\includegraphics[height=5.7cm]{fig3b.ps}
\caption{NIR and radio images of the Eye. ({\em left}) $2.16\mu$m continuum image of the field centred on
the Eye, overlaid with the Br$\gamma$ line image. 
Contours are at 3, 5, 7, 9, 12, 15 and $16 \times 10^{-19}$ W\,m$^{-2}$\,arcsec$^{-2}$.
({\em right}) 6 cm VLA image overlaid with contours of 20 cm VLA
image.  Contours are at 8, 26, 52, 104, 156, 208 and 
$258 \times 10^{-3}$\,Jy\,beam$^{-1}$.}
\end{figure}

Using the 3.8-m UK Infrared Telescope in conjuction with the CGS4
spectrometer we found 2.16$\mu$m Br$\gamma$ emission towards
the Eye. The emission peaks at a velocity of $\sim -200$\kms, which
drastically differs from the velocity of the molecular gas associated
with the Tornado. The velocity of the Br$\gamma$ emission indicates the distance 
to the Eye is 8.5 kpc, which makes the Eye foreground to the Tornado
nebula. A similar velocity towards the Eye has also been measured
using the H92$\alpha$  radio recombination  line (Brogan \& Goss 2003).

The Eye is resolved by the NIR and radio measurements (see Figure
3) as a compact  H{\sc ii} region, and therefore must 
be undergoing massive star formation.
It consists of four knots of emission, each about 1.5$\arcsec$ across and of
similar brightness,  placed symmetrically about the perimeter of a
circle 6$\arcsec$ across.  There are faint extensions extending 
$\sim$2$\arcsec$ to the south and to the west in the Br$\gamma$ image, but
no emission from its centre.

A fit to flux measurements from our NIR and other IR data obtained 
from {\em Midcourse Space Explorer (MSX)} and {\em Infrared Astronomy
Satellite (IRAS)} is consistent with the Eye being a warm ($\sim
190$\,K), unresolved ($\sim 0.05$\arcsec) blackbody source at
the core of an extended ($\sim 5.5$\arcsec), cold ($\sim 35$\,K) greybody.
 The best fit value of this two-component greybody gives an angular
size for the Eye which is similar to the size derived from the NIR and
radio images. The Eye's integrated infrared luminosity of
$\sim 2\times 10^{4}$\lsol\ suggests it harbors a massive ($\sim$12\msol)
protostellar source, perhaps a B0 star (see Burton et al. 2003 for more details).


\begin{thebibliography}{10}

\bibitem{bib1} Becker, R. H. \& Helfand, D. J., Nature, \textbf{313}, 115 (1985)

\bibitem{bib2} Brogan \& Goss, AJ, \textbf{125}, 272 (2003)

\bibitem{bib3} Burton, M. G., Lazendic, J. S., Yusef-Zadeh, F. \&
Wardle, M., to be submitted to MNRAS (2003)  

\bibitem{bib4} Caswell, J. L.; Haynes, R. F.; Milne, D. K.; Wellington, K. J., MNRAS,
\textbf{190}, 881 (1980)

\bibitem{bib5} Frail, D. A., Goss, W. M., Reynoso, E. M., Giacani, E. B., Green,
A. J. \&  Otrupcek, R., AJ \textbf{111} 1651 (1996)

\bibitem{bib6} Koralesky, B., Frail, D. A., Goss, W. M., Claussen, M. J. \& Green,
A. J., \textbf{116}, 1323 (1998)

\bibitem{bib7} Kundu, M. R., Velusamy, T. \& Hardee, P. E., AJ, \textbf{79}, 132 (1974)

\bibitem{bib8} Lazendic, J. S., Wardle, M., Burton, M. G., Yusef-Zadeh, F., Whiteoak,
J. B., Green, A. J. \&  Ashley, M. C. B., MNRAS, \textbf{331}, 537 (2002a)

\bibitem{bib9} Lazendic, J. S., Wardle, M., Green, A. J., Whiteoak, J. B. \& Burton,
M. G., Neutron Stars in Supernova Remnants,  Eds. P. O. Slane and
B. M. Gaensler, p.339 (2002b)

\bibitem{bib10} Lazendic, J. S., Wardle, M., Burton, M. G.,
Yusef-Zadeh, F., Whiteoak, J. B., Green, to be submitted to MNRAS (2003)

\bibitem{bib11} Lockett, P., Gauthier, E. \& Elitzur, M., ApJ,
\textbf{511}, 235 (1999)

\bibitem{bib12} Shaver, P. A., Salter, C. J., Patnaik, A. R., van
Gorkom, J. H. \& Hunt, G. C., Nature, \textbf{313}, 113 (1985a)

\bibitem{bib13} Shaver, P. A., Pottasch, S. R., Salter, C. J., Patnaik, A. R., van
Gorkom, J. H. \& Hunt, G. C., A\&A, \textbf{147}, L23 (1985b)

\end{thebibliography}
\end{document}